\documentclass[final,1p,times]{elsarticle}

\usepackage{lineno}
\usepackage{amsfonts}
\usepackage{graphicx}
\usepackage{subfigure}
\usepackage{amsmath}
\usepackage{algorithm2e}
\usepackage{float}

\modulolinenumbers[5]

\journal{Journal of \LaTeX\ Templates}

\begin{document}

\begin{frontmatter}

\title{Adversarial adaptive 1-D convolutional neural networks for bearing fault diagnosis under varying working condition}

\author[mymainaddress,mysecondaryaddress]{Bo Zhang}

\author[mymainaddress]{Wei Li}

\author[mythirdaddress]{Jie Hao}

\author[myforthaddress]{Xiao-Li Li}

\author[mymainaddress]{Meng Zhang}

\address[mymainaddress]{School of Mechatronic Engineering, China University of Mining and Technology Xuzhou 221116, People’s Republic of China}
\address[mysecondaryaddress]{School of Computer Science And Technology, China University of Mining and Technology Xuzhou 221116, People’s Republic of China}
\address[mythirdaddress]{School of Medicine Information, Xuzhou Medical University, Xuzhou 221004, People’s Republic of China}
\address[myforthaddress]{Institute for Infocomm Research (I$^2$R), A$^*$STAR, 1 Fusionopolis Way \#21-01 Connexis, 138632, Singapore}

\begin{abstract}
Traditional intelligent fault diagnosis of rolling bearings work well only under a common assumption that the labeled training data (source domain) and unlabeled testing data (target domain) are drawn from the same distribution. 
However, in many real-world applications, this assumption does not hold, especially when the working condition varies.
In this paper, a new adversarial adaptive 1-D CNN called A2CNN is proposed to address this problem. 
A2CNN consists of four parts, namely, a source feature extractor, a target feature extractor, a label classifier and a domain discriminator. The layers between the source and target feature extractor are partially untied during the training stage to take both training efficiency and domain adaptation into consideration. 
Experiments show that A2CNN has strong fault-discriminative and domain-invariant capacity, and therefore can achieve high accuracy under different working conditions.
We also visualize the learned features and the networks to explore the reasons behind the high performance of our proposed model.
\end{abstract}


\end{frontmatter}

\section{Introduction}
\label{sec:Introduction}

Machine health monitoring is of great importance in modern industry. 
Failure of these machines could cause great economical loss, and sometimes poses threats to the people who work with the machines. Therefore, in order to keep the industrial machines working properly and reliably, demand for better and more intelligent machine health monitoring technique has never ceased \cite{vachtsevanos2006intelligent,qiao2017adaptive}.
Rolling element bearings  are the most commonly used components in rotating machinery, and bearing faults may result in significant breakdowns, and even casualties \cite{Albrecht1987,Jardine2006}.
Therefore, effective fault diagnosis plays a highly significant role in increasing the safety and reliability of machinery and preventing possible damage \cite{Leiyaguo2011}.

In recent years, deep learning techniques have achieved huge success in computer vision \cite{he2017mask,chen2018deeplab} and speech recognition \cite{xiong2017microsoft,zhang2017very}. Some deep learning techniques have already found their way into machine health monitoring systems. 
For example, 
Jia et al. took the frequency spectra generated by fast Fourier transform (FFT) as the input of a stacked autoencoder (SAE) with three hidden layers for fault diagnosis of rotary machinery components \cite{JIA2016303}. 
Zhu et al. proposed a SAE model for hydraulic pump fault diagnosis that used frequency features generated by Fourier transform \cite{zhuhuijie2015fault}. 
Liu et al. used the normalized spectrum generated by Short-time Fourier transform (STFT) of sound signals as the input of a SAE model consisting of two layers. 
Moreover, multi-domain statistical features including time domain features, frequency domain features and time-frequency domain features were fed into the SAE model as a way of feature fusion \cite{liangguo2016multifeatures,verma2013intelligent}. 
There are also some researchers focusing on deep belief network (DBN) \cite{althobiani2014approach,gan2016construction,tao2016bearing}.
Convolutional neural networks (CNN) \cite{krizhevsky2012imagenet,lecun1998gradient} as one of the most popular deep learning networks, which have been successfully used in image recognition, is also used to realize fault diagnosis of mechanical parts.
Many CNN architectures were proposed, such as VGGNet \cite{simonyan2014very}, ResNet \cite{heKaiming2016deep} and Inception-v4 \cite{szegedy2017inception}, for 2-D image recognition. 
Also, CNN models for 1-D vibration signal were proposed. 
For example, 1D raw time vibration signals were used as the inputs of the CNN model for motor fault detection in \cite{Ince2016Real}, which successfully avoided the time-consuming feature extraction process. 
Guo et al. \cite{guoXiaojie2016hierarchical} proposed a hierarchical CNN consisting of two functional layers, where the first part is responsible for fault-type recognition and the other part is responsible for fault-size evaluation.

Most of the above proposed methods are only applicable to the situation that the data used to train classifier and the data for testing are under the same working condition, which means that these proposed methods work well only under a common assumption: the labeled training data (source domain) and unlabeled testing data (target domain) are drawn from the same distribution.
However, many real recognitions of bearing faults show this assumption does not hold, especially when the working condition varies. 
In this case, the labeled data obtained in one working condition may not follow the same distribution in another different working condition in real applications. 
When the distribution changes, most fault diagnosis models need to be rebuilt from scratch using newly recollected labeled training data. 
However, it is very expensive, if not impossible, to annotate huge amount of training data in the target domain to rebuild such new model. 
Meanwhile, large amounts of labeled training data in the source domain have not been fully utilized yet, which apparently waste huge resources and effort.
As one of the important research directions of transfer learning, domain adaptation (DA) typically aims at minimizing the differences between distributions of different domains in order to minimize the cross-domain prediction error by taking full advantage of information coming from both source and target domains.
Recently, DA has been introduced into the field of bearings fault diagnosis, such as \cite{shenFei2015bearing,lu2017deep,Zhang:2017aa,zhang2018deep}.
For instance, 
Zhang et al. \cite{Zhang:2017aa} took 1-D raw time vibration signal as the input of the CNN model, which realize fault diagnosis under different working loads. The domain adaptation capacity of this model originates from the method named Adaptive Batch Normalization (AdaBN).
Lu et al. \cite{lu2017deep} integrated the maximum mean discrepancy (MMD) as the regularization term into the objective function of DNN to reduce the differences between distributions cross domains.

In general, the main problem existing in domain adaptation is the divergence of distribution between the source domain and the target domain.
We need to learn a new feature representation, which should be fault-discriminative and simultaneously be domain-invariant.
The fault-discriminative ability refers that the learned feature representation should minimize the label classifier error, i.e., has a good ability to identify different faults.
The domain-invariant ability means that the learned feature representation should maximize the domain classification loss for all domains. 
That is to say, instances sampled from the source and target domains have similar distributions in the learned feature space.
As a result, a domain classifier can’t distinguish whether data come from the source domain or from the target domain.

In 2014, Goodfellow et al. \cite{goodfellow2014generative} proposed Generative Adversarial  Nets (GAN). 
GAN simultaneously train two models: a generative model $G$ captures the data distribution and a discriminative model $D$ estimates the probability that a sample came from the training data o generated by $G$. 
Inspired by GAN, we designed a adversarial adaptive model based on 1-D CNN named A2CNN, which simultaneously satisfied the above fault-discriminative and domain-invariant requirements. 
The details of the model will be shown in Section \ref{sec:method}. 
To our best knowledge, this is the first attempt for solving the domain adaptation issues in fault diagnosis by introducing adversarial network. 

The main contributions of this literature are summarized as follows.

1) We propose a novel adversarial adaptive CNN model which consists of four parts, namely, a source feature extractor, a target feature extractor, a label classifier and a domain discriminator.
During the training stage, the layers between the source and target feature extractor are partially untied to take both training efficiency and domain adaptation into consideration.

2) This proposed model has strong fault-discriminative and domain-invariant capacity, and therefore can achieve high accuracy under different working conditions. We visualize the feature maps learned by our model to explore the intrinsic mechanism of proposed model in fault diagnosis and domain adaptation. 

3) Besides the commonly used fault diagnostic \textit{accuracy}, we introduce two new evaluation indicators, \textit{precision} and \textit{recall}. Compared with \textit{accuracy}, \textit{precision} and \textit{recall} can evaluate the reliability of a model for certain type of fault recognition in more detail.

The rest of paper is organized as follows. 
In Section \ref{sec:preliminary}, some preliminary knowledge that will be used in our proposed framework is briefly reviewed. 
Section \ref{sec:method} introduced the construction of our proposed A2CNN. 
A series of experiments are conducted in Section \ref{sec:Experimental}. 
Finally, we conclude this paper in Section \ref{sec:conclusion}.

\section{Preliminary Knowledge}
\label{sec:preliminary}
The above CNNs for fault diagnosis mentioned in \ref{sec:Introduction} work well only under a common assumption: The training and test data is drawn from the same distribution.
However, vibration signals used for fault diagnosis usually show disobedience of the above assumption. In the running process of rotating machinery, because of complicated working conditions, the distributions of fault data under varying working condition are not consistent. For example, the training samples for building the classifier might be collected under the work condition without the motor load, nevertheless the actual application is to classify the defects from a bearing system under different motor load states. Although the categories of defects remain unchanged, the target data distribution changes with the motor load varies.

Our ultimate goal is to be able to predict labels given a sample from one working condition while the classifier is trained by the samples collected in another working condition.
Then, the problem above can be regarded as a domain adaptation problem, which is a realistic and challenging problem in fault diagnosis.
To solve this challenge, a domain adaption technique, would be needed to learn a discriminative classifier or other predictor in the presence of a "shift" between training and test distribution by taking full advantage of information coming from both source and target domains.

\subsection{Domain Adaptation}
According to the survey on domain adaptation (DA) for classification \cite{5288526}, a domain $\mathcal{D}$ consists of two components: a feature space $\mathcal{X}$ and a marginal probability distribution $P_X$, where $X \in \mathcal{X}$.
Give a specific domain, a task $\mathcal{T}$ consists of two components: a label space $\mathcal{Y}$ and a prediction function $f(X)$. From a probabilistic view point, $f(X)$ can be written as the conditional probability distribution $P_{Y|X}$.
Given a source domain $\mathcal{D}_S$ and a corresponding learning task $\mathcal{T}_S$, a target domain $\mathcal{D}_T$ and a corresponding learning task $\mathcal{T}_T$, domain adaptation aims to improve the learning of the target predictive function $f_T$ in $\mathcal{D}_T$ using the knowledge in $\mathcal{D}_S$ and $\mathcal{T}_S$, where $\mathcal{D}_S \neq \mathcal{D}_T$ and $\mathcal{T}_S=\mathcal{T}_T$, i.e., the tasks are the same but the domains are different.

In real world applications of fault diagnosis, the working conditions (e.g. motor load and speed) may change from time to time according to the requisite of the production.
As a kind of classification problem, the goal of intelligent fault diagnosis is to train classifier with samples collected and labeled in one working condition to be able to classify samples from another working condition.
Samples collected under different working conditions can be regarded as different domains.
Correspondingly, the fault diagnosis settings in domain adaptation situation are as follows:
\begin{itemize}
\item The feature spaces between domains are the same, $\mathcal{X}^S=\mathcal{X}^T$, e.g. the fast Fourier transform (FFT) spectrum amplitudes of raw vibration temporal signals.
\item The label spaces between domains are the same, $\mathcal{Y}^S=\mathcal{Y}^T=\{1,...,K\}$, where $K$ is the quantity of fault types.
\item $P^{S}_{XY}$ and $P^{T}_{XY}$ only differ in the marginal probability distribution of the input data, i.e., $P^{S}_{X} \neq P^{T}_{X}$, while $P^{S}_{Y|X}=P^{T}_{Y|X}$.
\end{itemize}
which is similar to the assumptions in covariate shift \cite{SHIMODAIRA2000227,Sugiyama:2008aa,Huang:2006:CSS:2976456.2976532} or sample selection bias \cite{Zadrozny:2004:LEC:1015330.1015425}.

\subsection{Domain Divergence Measure}
\label{sec_ddm}
The main problem existing in domain adaptation is the divergence of distribution between the target domain and source domain.
Ben-David et al. \cite{Ben-David2010,Ben-David:2006:ARD:2976456.2976474} defines a divergence measure $d_{H{\Delta}H}(S,T)$ between two domains $S$ and $T$, which is widely used in the theory of nonconservative domain adaptation.
Using this notion, they established a probabilistic bound on the performance $\epsilon_T(h)$ of some label classifier $h$ from $T$ evaluated on target domain given its performance $\epsilon_S(h)$ on the source domain. Formally,
\begin{equation}
  \label{eq_domaindiscrepancy}
  {\epsilon_T(h) \le \epsilon_S(h)+\frac12d_{H{\Delta}H}(S,T)+\lambda}
\end{equation}
where $\lambda$ is supposed to be a negligible term and dose not depend on classifier $h$.

Eq. \ref{eq_domaindiscrepancy} tells us that to adapt well, one has to learn a label classifier $h$ which works well on source domain while reducing the $d_{H{\Delta}H}(S,T)$ divergence between $S$ and $T$.
Estimating $d_{H{\Delta}H}(S,T)$ for a finite sample is exactly the problem of minimizing the empirical risk of a domain classifier $h_d$ that discriminates between instances drawn from $S$ and instances drawn from $T$, respectively pseudo-labeled with 0 and 1.
More specifically, it involves the following steps:

\begin{enumerate}
  \item Pseudo-labeling the source and target instances with 0 and 1, respectively.
  \item Randomly sampling two sets of instances as the training and testing set.
  \item Learning a domain classifier $h_d$ on the training set and verifying its performance on the testing set.
  \item Estimating the distance as $\hat{d}_{H{\Delta}H}(S,T)=1-2\epsilon(h_d)$, where $\epsilon(h_d)$ is the test error.
\end{enumerate}

It's obvious that if two domains perfectly overlap with each other, $\epsilon(h_d) \approx 0.5$, and $\hat{d}_{H{\Delta}H}(S,T) \approx 0$. On the contrary, if two domains are completely distinct from each other, $\epsilon(h_d) \approx 0$, and $\hat{d}_{H{\Delta}H}(S,T) \approx 1$. Therefore, $\hat{d}_{H{\Delta}H}(S,T) \in [0,1]$. The lower the value is, the smaller two domains divergence.

\subsection{Generative Adversarial Networks}
In 2014, Goodfellow et al. proposed a novel method named Generative Adversarial Networks (GAN).
A GAN consists of two part: a generator $G$ that synthesizes data whose distribution closely matches that of the real data, and a discriminator $D$ that estimates the probability that a sample came from the real data rather than $G$ \cite{DBLP:conf/nips/GoodfellowPMXWOCB14}.

Similar to Section \ref{sec_ddm}, real data are labeled with 0 and data generated by $G$ are labeled with 1.
The discriminator is trained to maximize the probability of assigning the correct label to both real samples and samples from $G$ \cite{DBLP:conf/nips/GoodfellowPMXWOCB14}.
While, the training procedure for $G$ is to maximize the probability of $D$ making mistake.
Therefore, the two models $G$ and $D$ formulated as a two-player minimax game, are trained simultaneously.
A unique solution exists with $G$ recovering the real data and $D$ is unable to distinguish
between real and generated samples, i.e. $D(x)=0.5$ everywhere \cite{DBLP:conf/nips/GoodfellowPMXWOCB14}.

By comparing DA, domain divergence measure $d_{H{\Delta}H}$ and GAN, we observe that they have a similar objective, that is, finding a feature representation that the data drawn from different distributions or different domains have the same distribution and perfectly overlap with each other after mapping to the learned feature space.
As a result, a domain discriminator (resp. classifier) can’t distinguish between the real (resp. the source domain) data and the generated (resp. the target domain) data.

\section{Proposed adversarial adaptive 1-D CNN}
\label{sec:method}

\subsection{Problem Formalization}

Let the labeled source domain data as $D_{S}=\{(x^{i}_S,y^{i}_S)\}|_{i=1}^{N_S}$, where $x^{i}_S \in \mathcal{R}^{m \times 1}$ is the data instance and $y^{i}_S \in \{1,...,K\}$ is the corresponding class label.
While, $D_{T}=\{(x^{i}_T)\}|_{i=1}^{N_T}$ is the unlabeled target domain data.
Here, $N_S$ and $N_T$ are the numbers of instances in $D_{S}$ and $D_{T}$.
In addition, each data instance is pseudo-labeled with a domain label $d \in \{0,1\}$ respectively, which indicates whether the instance comes from the source domain ($d = 0$) or from the target domain ($d = 1$).

\begin{figure*}[!ht]
\centering
\includegraphics[width=\textwidth]{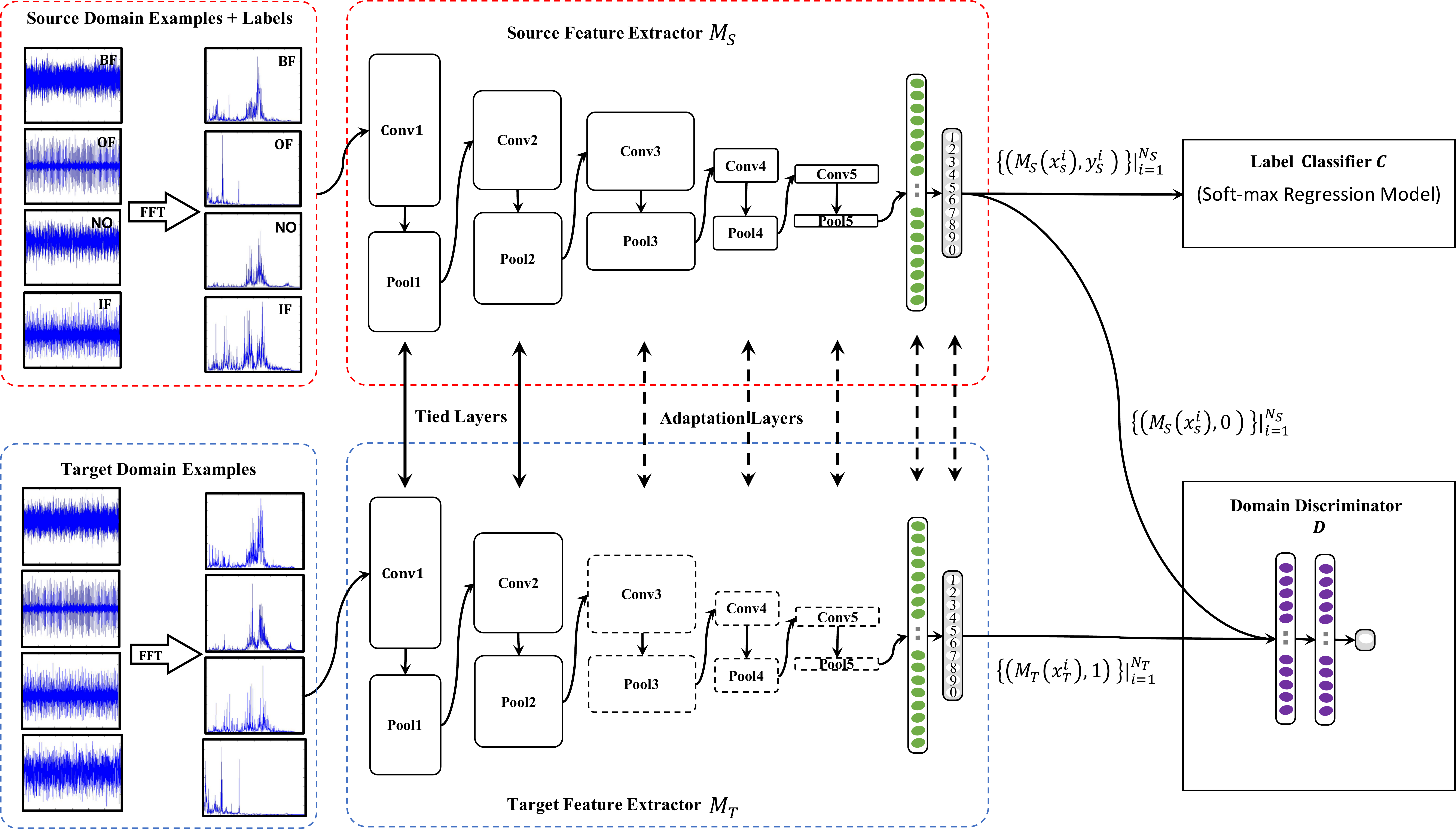}
\caption{The proposed Adversarial Adaptive 1-D CNN (A2CNN) includes a source feature extractor $M_S$,  a target feature extractor $M_T$, a label classifier $\mathcal{C}$ and a domain discriminator. Solid lines indicate tied layers, and Dashed lines indicate adaptive layers.}
\label{Fig_A2CNN}
\end{figure*}

The overall framework of the proposed Adversarial Adaptive 1-D CNN (A2CNN) is shown in Figure \ref{Fig_A2CNN}.
It includes a source feature extractor $M_S$, a target feature extractor $M_T$, a label classifier $C$ and a domain discriminator $D$, which together form a deep feed-forward architecture that maps each input sample $x^{i}_S$ (resp. $x^{i}_T$) to a $K$-dimensional feature vector $M_S(x^{i}_S)$ (resp. $M_T(x^{i}_T)$) ($K$ equals to the number of class label) and predicts its class label $y \in \{1,...,K\}$ and its domain label $d \in \{0,1\}$.

Compared with the traditional deep domain adaptation models, the proposed framework is more like an adversarial learning framework similar to GAN.
The parameters of $\mathcal{C}$ and $D$ should be optimized to minimizes the label prediction loss $\epsilon_S(\mathcal{C})$ (for the labeled source domain) and the domain classification loss $\epsilon(D)$ (for all domains).
And the parameters of the feature extractor $M_S$ and $M_T$ should be discriminative to minimize $\epsilon_S(\mathcal{C})$ and domain-invariant to maximize $\epsilon(D)$.

\paragraph{\textbf{Source feature extractor $M_S$}}
As shown in Figure \ref{Fig_A2CNN}, we compose the source feature extractor $M_S$ from five 1-D convolutional layers and two fully-connected layers.
The input of the first convolution layer (i.e. 'Conv1') is the fast Fourier transform (FFT) spectrum amplitudes of vibration signals, which is the most widely used approach of bearing defect detection.
The last fully-connected layer (i.e. 'FC2') is called label layer \cite{Zhuang:2015:SRL:2832747.2832823} with an output of $K$ neurons (equals to the number of class label), which is fed to label classifier $\mathcal{C}$ which estimate the posterior probability of each class.
It is common to add a pooling layer after each convolution layer in the CNN architecture separately.
It functions as a down-sampling operation which results in a reduced-resolution output feature map, which is robust to small variations in the location of features in the previous layer.
The most commonly used pooling layer is max-pooling layer, which performs the local max operation over the input features.
The main difference between the traditional 2-D and the 1-D CNN is the usage of 1-D arrays instead of 2-D matrices for both feature maps and filter kernels.
In order to capture the useful information in the intermediate and low frequency bands, the wide kernels should be used in the first convolutional layer which can better suppress high frequency noise\cite{Zhang:2017aa}. The following convolutional kernels are small (specifically, $3 \times 1$) which make the networks deeper to acquire good representations of the input signals and improve the performance of the network.

\paragraph{\textbf{Label classifier $\mathcal{C}$}}
For an source domain instance $x^{i}_S$, the output feature vector $M_S(x^{i}_S) \in \mathcal{R}^{K \times 1}$ mapped by the source feature extractor $M_S$ is the input of the label classifier $\mathcal{C}$.
Here, the soft-max regression model \cite{JSSv033i01} is used as the label classifier on source domain to incorporate label information.
The soft-max regression model is a generalization of the logistic regression model for multi-class classification problems.
We can estimate the probabilities of each class that $x^{i}_S$ belongs to as follows,
\begin{equation}
\label{equ:softmax}
\mathcal{C}\left(M_S(x^{i}_S)\right)=\begin{bmatrix} p\left(y=1|x^{i}_S\right) \\ p\left(y=2|x^{i}_S\right) \\ \vdots \\ p\left(y=K|x^{i}_S\right) \end{bmatrix}=\frac1{\sum_{j=1}^K{e^{u_j}}}\begin{bmatrix} e^{u_1} \\ e^{u_2} \\ \vdots \\ e^{u_K} \end{bmatrix},
\end{equation}
where $u_j=M_S(x^{i}_S)_j$ is the $j$-th value of $M_S(x^{i}_S)$,  $\sum_{j=1}^K{e^{u_j}}$ is a normalized term, and $p\left(y=j|M_S(x^{i}_S)\right)$ represent the distribution of the class $j \in \{1,2,...,K\}$ given the input $M_S(x^{i}_S)$.
Give the source domain data $D_S$, the parameters of the source feature extractor $M_S$ can be derived by maximizing the following supervised loss,
\begin{equation}
\label{equ:euq_L_cls}
\mathop{\max}_{M_S} L_{cls}(D_S)= \frac1{N_S}\sum_{i=1}^{N_S}\sum_{j=1}^K 1\{y^i_S=j\}\left[log\mathcal{C}\left(M_S(x^{i}_S)\right)\right].
\end{equation}
where $1\{y^i_S=j\}$ is an indicator function, whose value is 1 if $y^i_S=j$, otherwise is 0.

\paragraph{\textbf{Target feature extractor $M_T$}}
Based on "How to parametrize the target feature extractor", the approaches of most published adversarial adaptation works can be summarized into two categories: symmetric transformation and asymmetric transformation.
For many prior symmetric transformation methods \cite{conf/icml/GaninL15,DBLP:journals/corr/TzengHDS15}, all layers are constrained, thus enforcing exact source and target mapping consistency.
Although learning a symmetric transformation can reduce the number of parameters in the model, this may make the optimization poorly conditioned, since the same network must handle samples from two separate domains \cite{Tzeng2017}.
The intuitive idea behind the asymmetric transformation is to constrain a subset of the layers.
Rozantsev et al. \cite{DBLP:journals/corr/RozantsevSF16} showed that partially shared weights can lead to effective adaptation in both supervised and unsupervised settings.
We choose to learn parameters of the target feature extractor $M_T$ by partially untying layers between source and target mappings.
As shown in Figure \ref{Fig_A2CNN}, solid lines indicate tied layers, and dashed lines indicate adaptive layers.
Given that the target domain is unlabeled, we initialize the parameters of the target feature extractor $M_T$ with the source feature extractor $M_S$.

\paragraph{\textbf{Domain discriminator $D$}}
For an instance $x^{i}_S$ (resp. $x^{i}_T$), the output feature vector $M_S(x^{i}_S)$ (resp. $M_T(x^{i}_T)$) mapped by the feature extractor $M_S$ (resp. $M_T$), pseudo-labeled with a domain label $d^{i}=0$ (resp. $d^{i}=1$), is the input of the domain discriminator $D$.
The domain discriminator $D$ is a multi-layer perceptron (MLP), which is composed of several fully-connected layers (e.g. $input \to 500 \to  500 \to  1$).
The domain discriminator is in the binary classification setting. With the logistic regression model, its loss takes the form below,
\begin{equation}
\label{equ:euq_L_adv_D}
\begin{split}
\mathop{\max}_{D} L_{adv_D}=\frac1{N_S}\sum_{i=1}^{N_S}\left[logD(M_S(x^{i}_S))\right]+\frac1{N_T}\sum_{i=1}^{N_T}\left[log(1-D(M_T(x^{i}_T))\right].
\end{split}
\end{equation}

In order to obtain domain-invariant features, we seek the parameters of the target feature extractor $M_T$ to fool the domain discriminator $D$ by maximizing the following loss function $L_{adv_{M_T}}$ with inverted domain labels \cite{DBLP:conf/nips/GoodfellowPMXWOCB14},
\begin{equation}
\label{equ:euq_L_adv_M}
\mathop{\max}_{M_T} L_{adv_{M_T}}=\frac1{N_T}\sum_{i=1}^{N_T}\left[logD(M_T(x^{i}_T))\right].
\end{equation}

\subsection{Model Learning}

We have used two training steps to enhance the domain adaptation ability of our model.
The details of the proposed algorithm is summarized in Algorithm \ref{train}.

\begin{algorithm}
\caption{Adversarial Adaptive CNN (A2CNN)}\label{train}
 \DontPrintSemicolon
  \SetKwFunction{FPreTrain}{Pretrain}
  \SetKwFunction{FFineTune}{Finetune}
  \SetKwProg{Fn}{Function}{:}{end}
  \Fn{\FPreTrain{}}
  {
	\KwData{
	Given one source domain $D_{S}=\{(x^{i}_S,y^{i}_S)\}|_{i=1}^{N_S}$.
	}
	\KwResult{
	The parameters in the source feature extractor $M_S$.
	}
	\Begin
	{
	  \For{number of training iterations}
	  {
		Sample minibatch of $m$ instances $\{(x^{1}_S,y^{1}_S),...,(x^{m}_S,y^{m}_S)\}$ from the source domain $D_S$;
		
		Update the domain discriminator $M_S$ by ascending its stochastic gradient;
		$
		\bigtriangledown_{M_S}\frac1{m}\sum_{i=1}^{m}\sum_{j=1}^K 1\{y^i_S=j\}\left[log\mathcal{C}\left(M_S(x^{i}_S)\right)\right].
		$
	  }
	}
  }
  \Fn{\FFineTune{}}
  {
	\KwData{
	
	Given one source domain $D_{S}=\{(x^{i}_S,y^{i}_S)\}|_{i=1}^{N_S}$, and one target domain $D_{T}=\{(x^{i}_T)\}|_{i=1}^{N_T}$. 

	The parameters in the source feature extractor $M_S$. 

	The number of adaptive layers, $l$. 

	The number of steps to apply to the discriminator, $k$.
	}
	\KwResult{
	Results of the feature vector mapped by $M_T$, i.e., $\{M_T(x^{i}_T)\}|_{i=1}^{N_T}$.
	}
	\Begin
	{
	   Initialize the parameters of the target feature extractor $M_T$ with the source feature extractor $M_S$.

	  \For{number of training iterations}
	  {
		  \For{$i \leftarrow 1$ \KwTo $k$}
		  {
		  	Sample minibatch of $m$ instances $\{x^{1}_S,...,x^{m}_S\}$ from the source domain $D_{S}$.

			Sample minibatch of $m$ instances $\{x^{1}_T,...,x^{m}_T\}$ from the target domain $D_{T}$.

			Update the domain discriminator $D$ by ascending its stochastic gradient:
			$
			\bigtriangledown_D\frac1{m}\sum_{i=1}^{m}\left[logD(M_S(x^{i}_S))+log(1-D(M_T(x^{i}_T)))\right].
			$
		  }
		  Sample minibatch of $m$ instances $\{x^{1}_T,...,x^{m}_T\}$ from the target domain $D_{T}$.

		  Update the final $l$ adaptive layers of the target feature extractor $M_T$ by ascending its stochastic gradient:
		  $\bigtriangledown_{M_T}\frac1{m}\sum_{i=1}^{m}\left[logD(M_T(x^{i}_T))\right].$
	  }
	  \For{$j \leftarrow 1$ \KwTo $N_T$}
	  {
		  Computing the $M_T(x^{i}_T)$.
	  }
	}
  }
\end{algorithm}

\begin{enumerate}
\item \textbf{Pre-train}. Train the source feature extractor $M_S$ with labeled source training examples.
\item \textbf{Adversarial adaptive fine tune}.Initialize the parameters of the target feature extractor $M_T$ with the trained source feature extractor $M_S$ and learn a target feature extractor $M_T$ such that a domain discriminator $D$ can not predict the domain label of mapped source and target examples reliably.
\end{enumerate}

\subsection{Classifier Construction}
After all the parameters are learned, we can construct a classifier for the target domain by directly using the output of the last fully connected layer (i.e. 'FC2') of the target feature extractor $M_T$.
That is, for any instance $x^i_T$ in the target domain, the output of the target feature extractor $M_T(x^i_T)$ can computer the probability of instance $x^i_T$ belonging to a label $j \in \{1,...,K\}$ using Eq. \ref{equ:softmax}.
We choose the maximum probability using Eq. \ref{equ:euq_Classifier}. and the corresponding label as the prediction,
\begin{equation}
\label{equ:euq_Classifier}
y^i_T=\mathop{\max}_{j}\frac{e^{u_j}}{\sum_{l=1}^K{e^{u_l}}}, \ with \ u_j=M_T(x^{i}_T)_j.
\end{equation}

\section{Experimental analysis of proposed A2CNN model}
\label{sec:Experimental}

In real world applications, data under different load condition usually draw from different distribution. 
So it is significant to use unlabeled data under any load condition to rebuilt the classifier trained with samples collected in one load condition. 
In the reminder of this section, Case Western Reserve University (CWRU) bearing database is used to investigate how well the proposed A2CNN method performs under this scenario.

\subsection{Datasets and Preprocessing}

The test-bed in CWRU Bearing Data Center is composed of a driving motor, a two hp motor for loading, a torque sensor/encoder, a power meter, accelerometers and electronic control unit. 
The test bearings locate in the motor shaft. 
Subjected to electro-sparking, inner-race faults (IF), outer-race faults (OF) and ball fault (BF) with different sizes (0.007in, 0.014in, 0.021in and 0.028in) are introduced into the drive-end bearing of motor. 
The vibration signals are sampled by the accelerometers attached to the rack with magnetic bases under the sampling frequency of 12kHz. 
The experimental scheme simulates three working conditions with different motor load and rotating speed, i.e., Load1 = 1hp/1772rpm, Load2 = 2hp/1750rpm and Load3 = 3hp/1730rpm. 
The vibration signals of normal bearings (NO) under each working condition are also gathered.

In this paper, a vibration signal with length 4096 is randomly selected from raw vibration signal. 
Then, fast Fourier transform (FFT) is implemented on each signal and the 4096 Fourier coefficients are generated. 
Since the coefficients are symmetric, the first 2048 coefficients are used in each sample.
The samples collected from the above three different conditions form three domains, namely A, B and C, respectively.
There are ten classes under each working condition, including nine kinds of faults and a normal state, and each class consists of 800 samples.
Therefore, each domain contains 8000 samples of ten classes collected from corresponding working condition. 
The statistics of all domains are described in Table \ref{Table_case_test_set}.

\begin{table}[H]
\scriptsize
\begin{center}
  \caption{Description of the CWRU dataset}
  \label{Table_case_test_set}
  \begin{tabular}{c|c|ccc|ccc|ccc|c}
    \hline
    Category labels& 1 & 2 & 3 & 4 & 5 & 6 & 7 & 8 & 9 & 10 &\\
    \hline
    Fault location& None & \multicolumn{3}{c|}{IF} & \multicolumn{3}{c|}{BF} & \multicolumn{3}{c|}{OF} & Load\\
    \hline
    Fault diameter (in.)& 0 & 0.007 & 0.014 & 0.021 & 0.007 & 0.014 & 0.021 & 0.007 & 0.014 & 0.021 &\\
    \hline
    Domain A& 800 & 800 & 800 & 800 & 800 & 800 & 800 & 800 & 800 & 800 & 1\\
    Domain B& 800 & 800 & 800 & 800 & 800 & 800 & 800 & 800 & 800 & 800 & 2\\
    Domain C& 800 & 800 & 800 & 800 & 800 & 800 & 800 & 800 & 800 & 800 & 3\\
    \hline
    \#Features& 2048 & 2048 & 2048 & 2048 & 2048 & 2048 & 2048 & 2048 & 2048 & 2048 &\\
    \hline
  \end{tabular}
  \end{center}
\end{table}

To construct domain adaptation problems, we randomly choose two from the three domains, where one is considered as the source domain and the other is considered as the target domain.
Therefore, we construct six ($P^2_3$) domain adaptation problems. Take the domain adaption task $A \to B$ as an example. The examples of domain A are used as the source domain data $D_S$, and the examples of domain B are used as the target domain data $D_T$.

\subsection{Experimental setup}

\subsubsection{Baseline Methods}

We compare our methods with the following baselines,

\begin{enumerate}
  \item Traditional SVM and Multi-layer Perceptron (MLP) which work with the data transformed by Fast Fourier transformation (FFT).
  \item The deep neural network (DNN) system with frequency features \cite{JIA2016303} proposed  by Lei et al. in 2016.
  This neural network consists of three hidden layers. The number of neurons in each layer is 1025, 500, 200, 100 and 10. 
  The input of the network is the normalized 1025 Fourier coefficients transformed from the raw temporal signals using FFT. 
  \item The Deep Convolution Neural Networks with Wide first-layer kernels (WDCNN) system \cite{Zhang:2017aa} proposed by Zhang et al. in 2017. 
  The WDCNN system works directly on raw temporal signals. It contains five convolutional layers and batch normalization layers.
  The domain adaptation capacity of this model originates in the domain adaptation method named Adaptive Batch Normalization (AdaBN). 
\end{enumerate}

\subsubsection{Parameters of the proposed A2CNN}

The feature extractor $M_S$ and $M_T$ used in experiments is composed of five convolutional layers and pooling layers followed by two fully-connected hidden layers. 
The pooling type is max pooling and the activation function is ReLU. 
The parameters of the convolutional and pooling layers are detailed in Table \ref{Table_a2cnn}.
In order to minimize the loss function, the Adam Stochastic optimization algorithm is applied to train our CNN model.
The final $l$ ($l \in [1, 7]$) layers of the target feature extractor $M_T$ is untied and used as adaptive layers.
The domain discriminator $D$ consists of three fully-connected layers. The number of neurons in each layer is 500, 500 and 1.
The experiments were implemented using Tensorflow toolbox of Google. 
\begin{table}[H]
\scriptsize
\begin{center}
  \caption{Details of the feature extractor $M_S$ and $M_T$ used in experiments.}
  \label{Table_a2cnn}
  \begin{tabular}{ccccccc}
    \hline
    No. & Layer type & Kernel & stride & Channel & Output & Padding\\
    \hline
    1 & Convolution1 & $32 \times 1$ & $2 \times 1$ & 8  & $1009 \times 8$ & Yes \\
	2 & Pooling1     & $2  \times 1$ & $2 \times 1$ & 8  & $504  \times 8$ & No  \\

	3 & Convolution2 & $16 \times 1$ & $2 \times 1$ & 16 & $245 \times 16$ & Yes \\
	4 & Pooling2     & $2  \times 1$ & $2 \times 1$ & 16 & $122 \times 16$ & No  \\

	5 & Convolution3 & $8  \times 1$ & $2 \times 1$ & 32 & $58  \times 32$ & Yes \\
	6 & Pooling3     & $2  \times 1$ & $2 \times 1$ & 32 & $29  \times 32$ & No  \\

	7 & Convolution4 & $8  \times 1$ & $2 \times 1$ & 32 & $11  \times 32$ & Yes \\
	8 & Pooling4     & $2  \times 1$ & $2 \times 1$ & 32 & $5   \times 32$  & No  \\

	9 & Convolution5& $3   \times 1$ & $2 \times 1$ & 64 & $2   \times 64$ & Yes \\
	10& Pooling5    & $2   \times 1$ & $2 \times 1$ & 64 & $1   \times 64$ & No  \\

	11& Fully-connected & 500 &  & 1 & $500$ \\
	12& Fully-connected & 10  &  & 1 & 10 \\
    \hline
  \end{tabular}
  \end{center}
\end{table}

By contrast, for an instance from the target domain $x^i_T$, in order to investigate the effectiveness of adversarial adaptation, we use the corresponding output of the well-trained source feature extractor $M_S(x^i_T)$ to computer the probability of the instance $x^i_T$ belonging to a label $j \in \{1,...,K\}$ using Eq. \ref{equ:softmax}, which is denoted as A2CNN$_S$.

\subsection{Accuracy across different domains}
\label{subsection:case1_result}

\begin{figure}
\centering
\includegraphics[width=\textwidth]{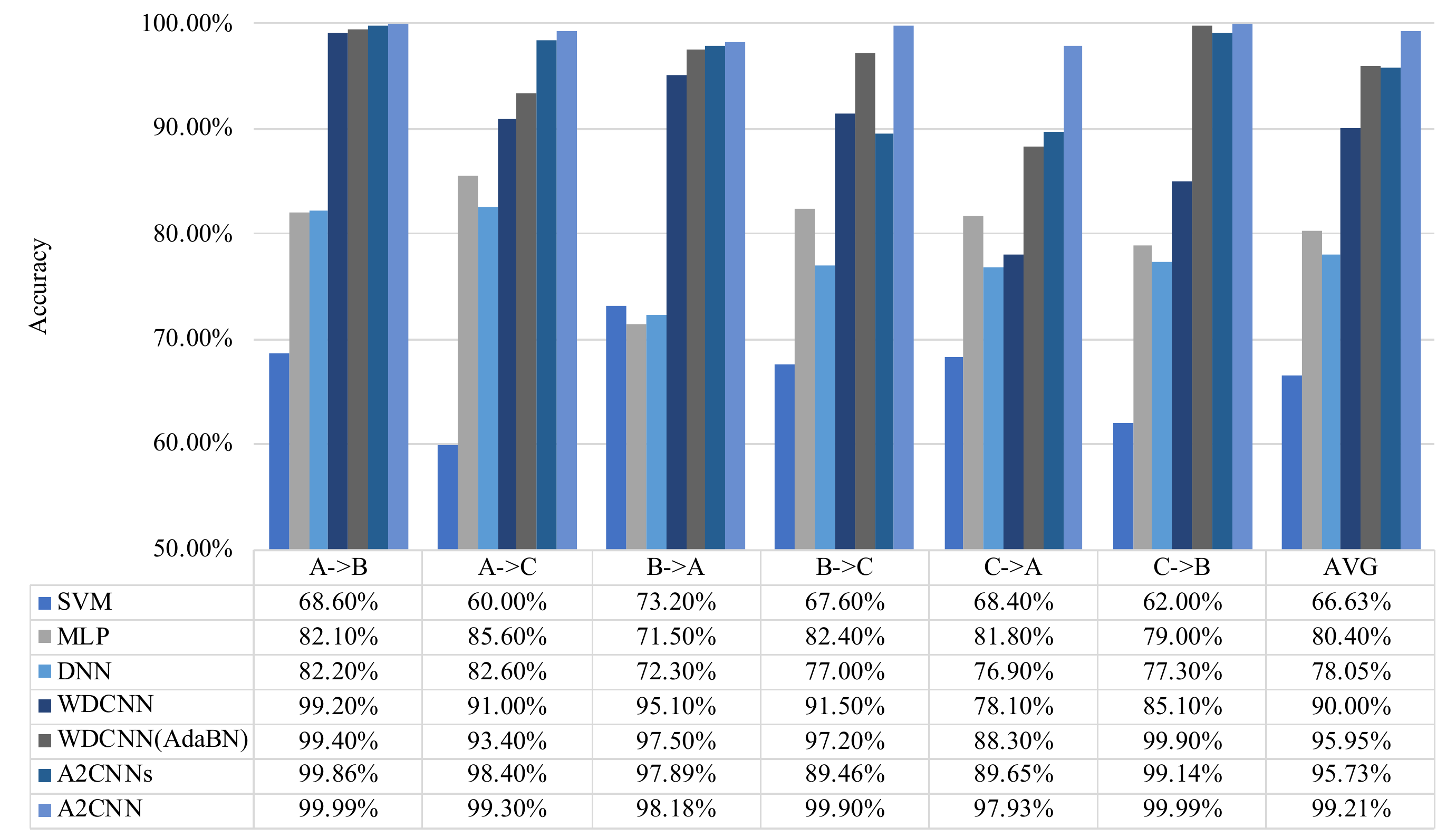}
\caption{Accuracy (\%) on six domain adaptation problems.}
\label{Fig_A2CNN_Results}
\end{figure}

As Figure \ref{Fig_A2CNN_Results} shows, SVM, MLP and DNN perform poorly in domain adaptation, with average accuracy in the six scenarios being around 65\%, 80\% and 80\%. 
Which prove that samples under different working conditions draw from the different distributions and models trained under one working condition is not suitable for fault classification under another working load condition. 

Compared with the WDCNN with AdaBN, which achieved the state-of-art domain adaptation ability, A2CNN which achieves 99.21\% accuracy in average is obviously greater than WDCNN (AdaBN) with average accuracy being 95.95\%. 
This result prove that the features learned by A2CNN are more domain invariant than the features learned by the other methods. 

In addition, by comparing A2CNN with A2CNN$_S$, we can find that in every scenario, the performance of A2CNN is superior to A2CNN$_S$. 
This means that the adversarial adaptation training can significantly improve the bearing fault diagnosis under varying working conditions.

It is also interesting that when adapting from Domain A to B, from B to A, from B to C, and from C to B, the fault diagnosis accuracy of the proposed A2CNN is only a bit better than WDCNN (AdaBN). 
However, when adapting from domain A to C and C to A, the proposed A2CNN is significantly better than the other methods. 
This result prove that A2CNN is good at solving the problem that the distributions of source domain and target domain are far different.

\subsection{Sensitivity Analysis of Faults}

For each type of fault detection, in order to further analyze the sensitivity of the proposed A2CNN model, we introduce two new evaluation indicators, i.e. \textit{precision} and \textit{recall}, which are widely used in pattern recognition, information retrieval and binary classification.

In the fault diagnosis context, the \textit{precision} and \textit{recall} for a kind of fault type $f$ can be calculated as below, 

\begin{equation}
\label{equ:euq_precision_recall}
\textit{precision}(f)=\frac{TP}{TP + FP},
\textit{recall}(f)=\frac{TP}{TP + FN},
\end{equation}
where \textit{true positives} ($TP$) means the number of faults correctly identified as $f$, \textit{false positives} ($FP$) means the number of faults incorrectly labeled as $f$ and \textit{false negatives} ($FN$) means the number of faults $f$ incorrectly \ labeled as not belonging to $f$.

A \textit{precision} score of 1.0 for a fault type $f$ means that every sample labeled as belonging to class $f$ does indeed belong to class $f$ (i.e. there is no false alarm), but it can't tell us about the number of samples from class $f$ that were not labeled correctly (i.e. how many failures are missing?).

Whereas a \textit{recall} of 1.0 means that every item from a fault type $f$ was labeled as belonging to class $f$ (i.e. there is no missing alarm), but says nothing about how many other items were incorrectly also labeled as belonging to class $f$ (i.e. how many false alarms are there?).

The \textit{precision} and \textit{recall} of every class processed by A2CNN and A2CNN$_S$ are detailed in Table \ref{Table_precision} and Table \ref{Table_recall}. 

In Table \ref{Table_precision}, for the 3rd kinds of fault (i.e. IF with fault size being 0.014 in.), A2CNN$_S$ has low \textit{precision} when adapting from domain B to C and from C to A, which are 49.63\% and 57.55\% respectively.
This means that about half of that kind of fault alarms are unreliable.

Meanwhile, in Table \ref{Table_recall}, for the 2nd kinds of fault (i.e. IF with fault size being 0.007 in.), A2CNN$_S$ has very low \textit{recall} when adapting from domain B to C and from C to A, which are 1.13\% and 26.75\% respectively.
This means that about a large number of that kind of failures are not detected.

In general, the \textit{precision} and \textit{recall} of A2CNN are higher than that of A2CNN$_S$, which implies that A2CNN has fewer false alarms (i.e. high \textit{precision} score ) and missed alarms (i.e. high \textit{recall} score ). 
We can find that A2CNN can make almost all class classified into right class, except BF with fault size being 0.014 in and BF with fault size being 0.021 in. 
This result shows that after adversarial fine tuning, the classification performance on every class achieve remarkable improvement.

\begin{table}[H]
\scriptsize
\begin{center}
  \caption{\textit{precision} of the proposed A2CNN$_S$ and A2CNN on six domain adaptation problems.}
  \label{Table_precision}
  \begin{tabular}{c|c|ccc|ccc|ccc}
    \hline
    Fault location & None & \multicolumn{3}{c|}{IF} & \multicolumn{3}{c|}{BF} & \multicolumn{3}{c}{OF} \\
    \hline
    Fault diameter (in.)&  & 0.007 & 0.014 & 0.021 & 0.007 & 0.014 & 0.021 & 0.007 & 0.014 &0.021\\
    \hline
    Category labels& 1 & 2 & 3 & 4 & 5 & 6 & 7 & 8 & 9 & 10\\
    \hline
    \multicolumn{11}{l}{\textit{precision} of A2CNN$_S$} \\
    \hline
	A$\to$B & 100\% & 100\% & 100\% & 100\% & 100\% & 100\% & \textbf{98.64\%} & 100\% & 100\% & 100\% \\
	A$\to$C & \textbf{92.27\%} & 100\% & 100\% & 100\% & \textbf{99.75\%} & \textbf{92.82\%} & \textbf{99.46\%} & 100\% & 100\% & 100\% \\
	B$\to$A & 100\% & 100\% & 100\% & 100\% & \textbf{87.43\%} & 93.68\% & 100\% & 100\% & 100\% & 100\% \\ 
	B$\to$C & \textbf{96.74\%} & 100\% & \underline{\textbf{49.63\%}} & 100\% & 100\% & \textbf{99.49\%} & 100\% & 100\% & 100\% & 100\% \\ 
	C$\to$A & 100\% & 100\% & \underline{\textbf{57.55\%}} & 100\% & \textbf{83.33\%} & 95.40\% & 100\% & \textbf{95.12\%} & 100\% & 100\% \\ 
	C$\to$B & 100\% & 100\% & \textbf{93.13\%} & \textbf{99.88\%} & \textbf{98.89\%} & 100\% & 100\% & 100\% & 100\% & 100\% \\ 
    \hline
    \multicolumn{11}{l}{\textit{precision} of A2CNN} \\
    \hline
	A$\to$B & 100\% & 100\% & 100\% & 100\% & 100\% & 100\% & 99.88\% & 100\% & 100\% & 100\% \\ 
	A$\to$C & 93.46\% & 100\% & 100\% & 100\% & 100\% & 100\% & 100\% & 100\% & 100\% & 100\% \\ 
	B$\to$A & 100\% & 100\% & 100\% & 100\% & 90.70\% & \textbf{92.59\%} & 100\% & 100\% & 100\% & 100\% \\ 
	B$\to$C & 100\% & 100\% & 99.01\% & 100\% & 100\% & 100\% & 100\% & 100\% & 100\% & 100\% \\ 
	C$\to$A & 100\% & 100\% & 100\% & 100\% & 90.91\% & \textbf{90.40\%} & 100\% & 100\% & 100\% & 100\% \\ 
	C$\to$B & 100\% & 100\% & 100\% & 100\% & 99.88\% & 100\% & 100\% & 100\% & 100\% & 100\% \\ 
    \hline															
  \end{tabular}
  \end{center}
\end{table}

\begin{table}[H]
\scriptsize
\begin{center}
  \caption{\textit{recall} of the proposed A2CNN$_S$ and A2CNN on six domain adaptation problems.}
  \label{Table_recall}
  \begin{tabular}{c|c|ccc|ccc|ccc}
    \hline
    Fault location & None & \multicolumn{3}{c|}{IF} & \multicolumn{3}{c|}{BF} & \multicolumn{3}{c}{OF} \\
    \hline
    Fault diameter (in.)&  & 0.007 & 0.014 & 0.021 & 0.007 & 0.014 & 0.021 & 0.007 & 0.014 & 0.021\\
    \hline
    Category labels& 1 & 2 & 3 & 4 & 5 & 6 & 7 & 8 & 9 & 10\\
    \hline
    \multicolumn{11}{l}{\textit{recall} of A2CNN$_S$} \\
    \hline
	A$\to$B & 100\% & 100\% & 100\% & 100\% & \textbf{98.63\%} & 100\% & 100\% & 100\% & 100\% & 100\% \\
	A$\to$C & 100\% & 100\% & \textbf{99.75\%} & 100\% & \textbf{99.50\%} & \textbf{92.13\%} & \textbf{92.88\%} & 100\% & \textbf{99.50\%} & 100\% \\
	B$\to$A & 100\% & 100\% & 100\% & 100\% & 100\% & 100\% & \textbf{78.88\%} & 100\% & 100\% & 100\% \\
	B$\to$C & 100\% & \underline{\textbf{1.13\%}} & 100\% & 100\% & 100\% & \textbf{96.63\%} & \textbf{99.50\%} & 100\% & \textbf{97.38\%} & 100\% \\
	C$\to$A & 100\% & \underline{\textbf{26.75\%}} & 100\% & 100\% & 100\% & \textbf{96.00\%} & \textbf{73.75\%} & 100\% & 100\% & 100\% \\ 
	C$\to$B & 100\% & \textbf{92.63\%} & 100\% & 100\% & 100\% & \textbf{99.88\%} & \textbf{98.88\%} & 100\% & 100\% & 100\% \\
    \hline
    \multicolumn{11}{l}{\textit{recall} of A2CNN} \\
    \hline
	A$\to$B & 100\% & 100\% & 100\% & 100\% & 100\% & \textbf{99.88\%} & 100\% & 100\% & 100\% & 100\% \\
	A$\to$C & 100\% & 100\% & 100\% & 100\% & 100\% & 93.00\% & 100\% & 100\% & 100\% & 100\% \\
	B$\to$A & 100\% & 100\% & 100\% & 100\% & 100\% & 100\% & 81.75\% & 100\% & 100\% & 100\% \\
	B$\to$C & 100\% & 100\% & 100\% & 100\% & 100\% & 100\% & 100\% & 100\% & 99.00\% & 100\% \\
	C$\to$A & 100\% & 100\% & 100\% & 100\% & 100\% & 100\% & 79.38\% & 100\% & 100\% & 100\% \\
	C$\to$B & 100\% & 100\% & 100\% & 100\% & 100\% & 100\% & 99.88\% & 100\% & 100\% & 100\% \\ 
    \hline															
  \end{tabular}
  \end{center}
\end{table}

\subsection{Parameter Sensitivity}
\label{sec_parameter_sensitivity}
In this section, we investigate the influence of the parameter $l$, which represents the number of untied layers in the target feature extractor $M_T$ during the adversarial adaptive fine tune.
Given that the target feature extractor $M_T$ contains five convolutional layers and pooling layers and two fully-connected hidden layers, $l$ is selected from $\{1,...,7\}$ in our experiment. 
We use A2CNN$_l$ to denote the A2CNN model with the parameter $l$. 
For example, A2CNN$_1$ indicates that only the last fully-connected hidden layer is untied (i.e. FC2 in Figure \ref{Fig_A2CNN}), and A2CNN$_7$ means all the seven layers in $M_T$ are untied (i.e. from 'Conv1' to 'FC2' in Figure \ref{Fig_A2CNN}).

\begin{figure}[H]
\centering
\includegraphics[width=0.8\textwidth]{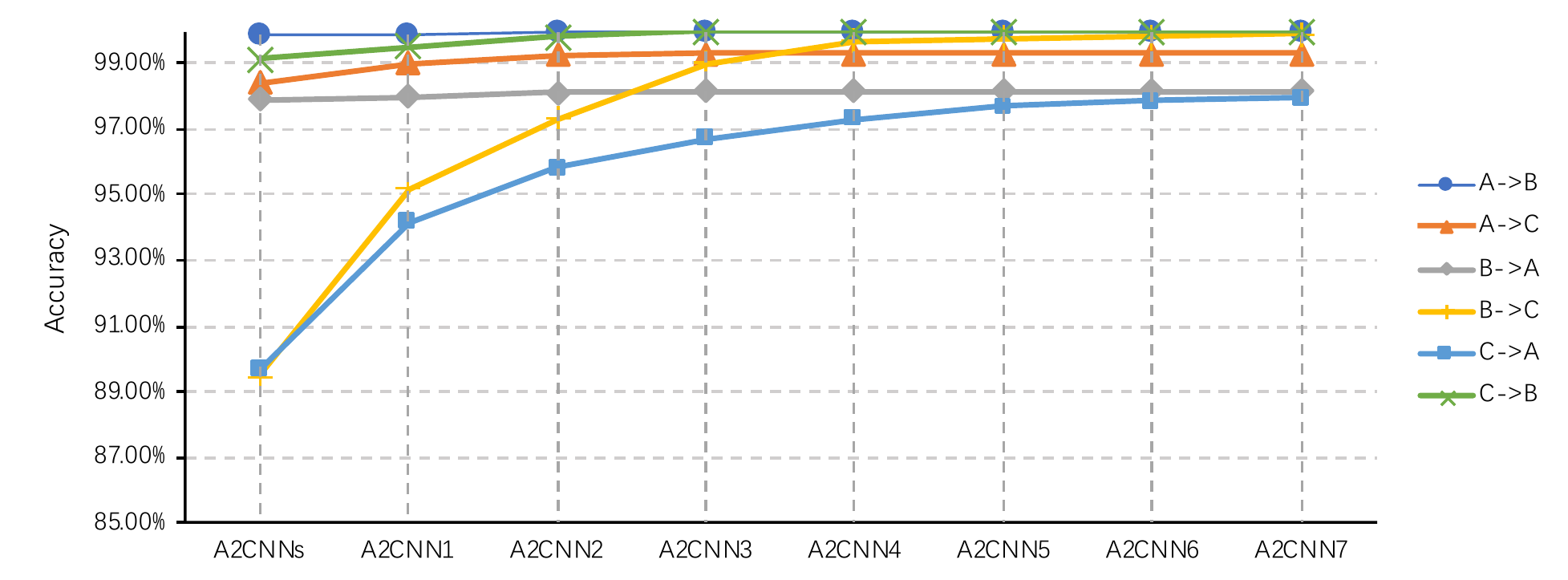}
\caption{The Parameter influence of the number of untied layers $l$ on A2CNN.}
\label{Fig_A2CNN_L}
\end{figure}

Figure \ref{Fig_A2CNN_L} reports the results. From the figure, we can generally observe that the more untied layers involved in adversarial adaptive fine tuning stage, the higher the accuracy of recognition. However, the sensitivity of different adaptive problems to parameters $l$ is different.

First of all, when adapting from domain A to B and from B to A, the enhancement of recognition accuracy is limited. We can use A2CNN$_S$ directly to achieve the accuracy of 99.86\% and 97.89\% respectively, which is only a little worse than A2CNN$_7$.
Moreover, in these two cases, the \textit{precision} and \textit{recall} of A2CNN$_S$ are also very similar to A2CNN$_7$. 
This  can be interpreted as that there is little difference in the distribution between domain A and domain B.

Then, for the domain adaptation from A to C and from C to B, we only need to untie the last  two fully-connected hidden layers (i.e. A2CNN$_2$) to achieve the same highest accuracy as A2CNN$_7$. 
However, the distribution differences between domains are not symmetrical.
By contrast, for the domain adaptation from C to A and from B to C, we have to respectively untie the last six (i.e. A2CNN$_6$) and four (i.e. A2CNN$_4$) layers to achieve the almost best results.

\subsection{Networks visualizations}
\label{subsection:Networks visualizations}

Generally, deep learning is an empirical success rather than a mathematical solution to the learning problem. 
In order to explain why the proposed A2CNN model can achieve such a great performance in bearing fault diagnosis under varying working conditions vividly, the features extracted by the $M_S$ and $M_T$ are visualized in this subsection. 

\textit{t-Distributed Stochastic Neighbor Embedding} (\textit{t-SNE}) is a technique for dimensionality reduction that is wildly used for the visualization of deep neural networks. 
The goal of \textit{t-SNE} is to take a set of points in a high-dimensional space and find a faithful representation of those points in a lower-dimensional space, typically the 2D plane. 
In this paper, \textit{t-SNE} is used to visualize the features extracted by A2CNN. 
For more details about \textit{t-SNE} one can refer to Ref.\cite{P2017Visualizing}.

\begin{figure}[H]
\centering
\includegraphics[width=\textwidth]{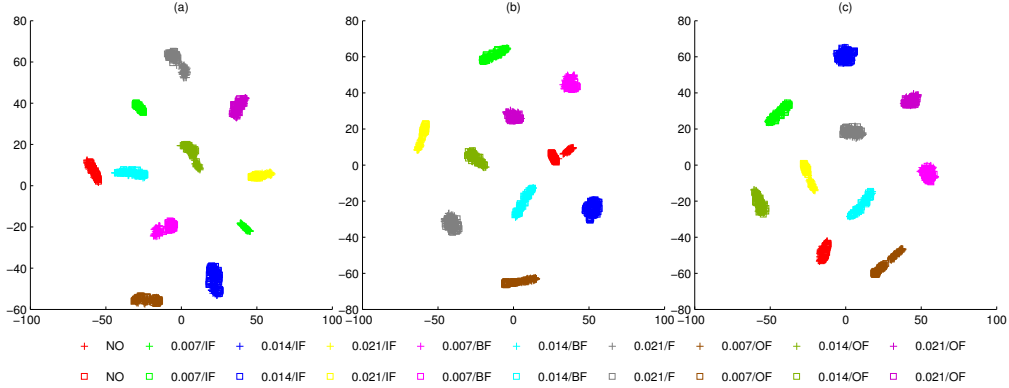}
\caption{
Visualization of the extracted features of samples collected from the source domain B and target domain C via \textit{t-SNE}.
Square symbols represent the features of the sample $\{x^i_S\}|^{N_S}_{i=1}$ collected from the source domain B. 
Cross symbols represent the features of the sample  $\{x^i_T\}|^{N_S}_{i=1}$  collected from the target domain C. 
Ten different kinds of faults are denoted by ten different colors respectively.
}
\label{Fig_A2CNN_tSNE_2}
\end{figure}

Take the domain adaption task $B \to C$ as an example, \textit{t-SNE} is used to visualize the high-dimensional features extracted by the source feature extractor $M_S$ and the target feature extractor $M_T$. 
The result is shown in Figure\ref{Fig_A2CNN_tSNE_2}.
In all subgraphs of Figure \ref{Fig_A2CNN_tSNE_2}, features of the source sample $\{x^i_S\}|^{N_S}_{i=1}$ are extracted by $M_S$, i.e., $M_S(x^i_S)|^{N_S}_{i=1}$. 
For the target sample, features extracted by $M_S$ (i.e., $M_S(x^i_T)|^{N_T}_{i=1}$) are shown in (a)
and 
features extracted by the fine-tuning $M_T$ after 1000 and 2000 iterations are shown in (b) and (c).
For convenience, these features are denoted by $M_S(x_S)$, $M_S(x_T)$ and $M^{it}_T(x_T)$, where the number of iterations $it$ is selected from $\{1000, 2000\}$.

There are some interesting observations as follows.
\begin{enumerate}
\item $M_S(x_S)$, $M_S(x_T)$ and $M^{it}_T(x_T)$ are classifiable when they are diagnosed separately. 
This illustrates that 1-D CNN used for $M_S$ and $M_T$ has a very strong ability to distinguish various rolling bearing faults and explains the reason why A2CNN$_S$ can even achieve such a good classification accuracy.

\item In Figure\ref{Fig_A2CNN_tSNE_2}(a), the distribution of fault '0.007/IF' is completely different between domains. This explains why A2CNN$_S$ has the very low recall of 1.13\% when adapting from domain B to C.

\item During the stage of fine tuning, with the increasing of the adversarial iterations, the distributions of features between $M_S(x_S)$ and $M^{it}_T(x_T)$ gradually become consistent. 
When the features $M^{1000}_T(x_T)$ and $M^{2000}_T(x_T)$ are applied to fault detection, the accuracies are 99.75\% and 99.90\% respectively. 
The observation indicates that the adversarial network is effective in improving the domain adaptation capacity of the A2CNN.
\end{enumerate}

\begin{figure}
\centering
\subfigure[$M_S(x^{0.007/IF}_C)$ and the corresponding soft-max result.]{
\begin{minipage}{\textwidth}
\includegraphics[width=\textwidth]{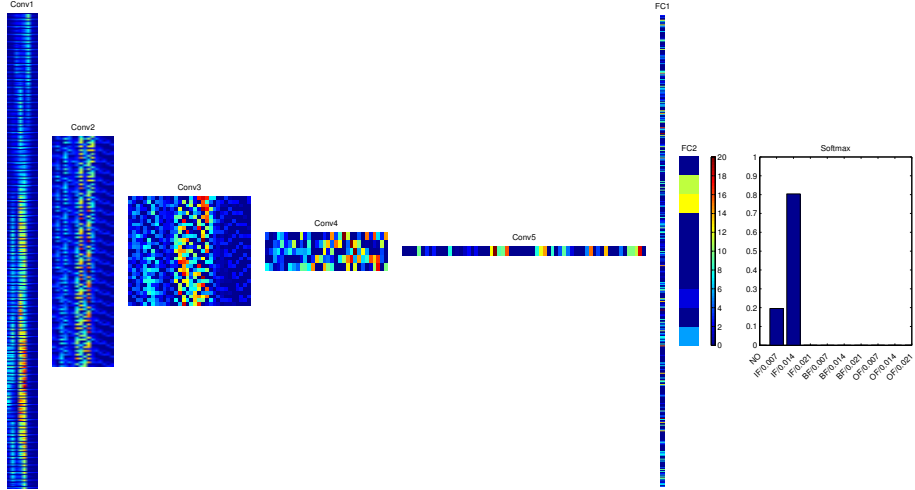}
\end{minipage}
}
\subfigure[$M_T(x^{0.007/IF}_C)$ and the corresponding soft-max result.]{
\begin{minipage}{\textwidth}
\includegraphics[width=\textwidth]{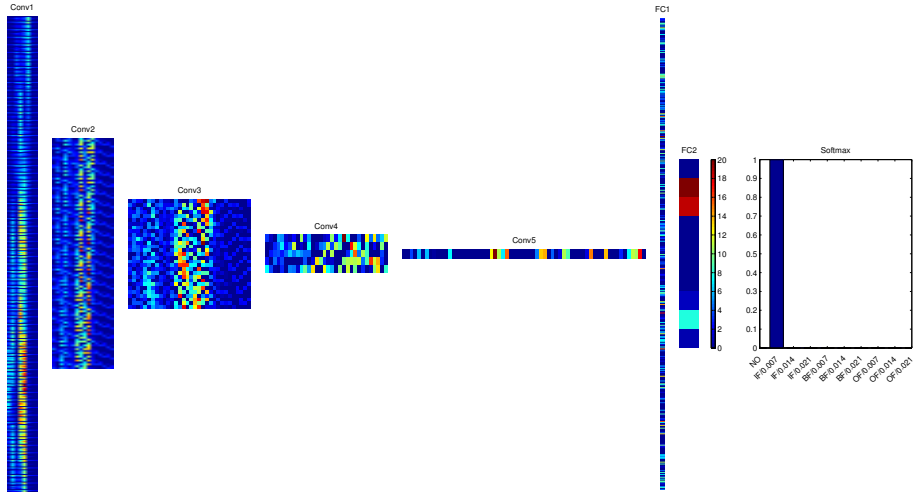}
\end{minipage}
}
\caption{
Visualization of all nodes in $M_S$, $M_T$ and the soft-max results of the label predictor $\mathcal{C}$. Domain B is the source domain and domain C is the target domain.
}
\label{Fig_A2CNN_activation}
\end{figure}

Finally, we visualize all nodes in the entire A2CNN model, including $M_S$, $M_T$ and the soft-max outputs of the label predictor $\mathcal{C}$. 
We randomly select a sample of the fault type '0.007/IF' from domain C, denoted by $x^{0.007/IF}_C$, as the input of the A2CNN model trained for adapting from domain B to domain C.
The visualized results are shown in Figure \ref{Fig_A2CNN_activation}(a) and Figure \ref{Fig_A2CNN_activation}(b).

From these visual results, we can find out that the output of the first three convolutional layers (i.e. 'Conv1', 'Conv2' and 'Conv3') are very similar.
Starting from the fourth layer convolution layer (i.e. 'Conv4'), the extracted features of $M_S(x^{0.007/IF}_C)$ and $M_T(x^{0.007/IF}_C)$ gradually change to some extent.
This observation is consistent with the result in the section \ref{sec_parameter_sensitivity}. 
That is, for the domain adaptation from B to C, we only have to untie last four layers starting from 'Conv4' to 'FC2' as the features extracted from the the first three convolutional layers are almost the same. 

The last fully-connected layer (i.e. 'FC2') has $K$ neurons (equals to the number of class label).
The output of 'FC2' is fed to label classifier $\mathcal{C}$ to estimate the posterior probability of each class using soft-max regression model.

According to the results of 'FC2' and 'Softmax' in Figure \ref{Fig_A2CNN_activation}(a), $x^{0.007/IF}_C$ is misdiagnosed as the fault type of '0.014/IF', based on the extracted features of $M_S(x^{0.007/IF}_C)$.
As a contrast, in Figure \ref{Fig_A2CNN_activation}(b), $x^{0.007/IF}_C$ is correctly identified as as the fault type of '0.007/IF', based on the extracted features of $M_T(x^{0.007/IF}_C)$.

\section{Conclusion}
\label{sec:conclusion}

This paper proposes a adversarial adaptive model based on 1-D CNN named A2CNN, to address the fault diagnosis problem under varying working condition. 
To our best knowledge, this is the first attempt for solving the domain adaptation issues in fault diagnosis by introducing adversarial network. 

A2CNN contains four parts, a source feature extractor, a target feature extractor, a label classifier and a domain discriminator.
In order to get the strong fault-discriminative and domain-invariant capacity, we adopt the training process similar to Generating Adversarial Network (GAN). 
First, in order to get fault-discriminative features, the source feature extractor is pre-trained with labeled source training examples to minimize the label classifier error.
Then, during the adversarial adaptive fine tuning stage, the target feature extractor is initialized and trained to maximize the domain classification loss, such that the domain discriminator can't predict the domain label of mapped source and target examples reliably.
That is to say, instances sampled from the source and target domains have similar distributions after mapping.
In addition, the layers between the source and target feature extractor during the training stage are partially untied to give consideration to the training efficiency and the domain adaptability.

Results in Section \ref{sec:Experimental} shows that, compared with the state-of-the-art domain adaptive model WDCNN (AdaBN), the proposed A2CNN achieve higher \textit{accuracy} under different working conditions. 
Besides the commonly used fault diagnostic \textit{accuracy}, we introduce two new evaluation indicators, \textit{precision} and \textit{recall}, to analyze the sensitivity of the proposed for each type of fault detection.
A \textit{precision} score of 1.0 for a fault type means that there is no false alarm.
Whereas a \textit{recall} of 1.0 means that there is no missing alarm.
Compared with \textit{accuracy}, \textit{precision} and \textit{recall} can evaluate the reliability of a model for certain type of fault recognition in more detail.
This result shows that after adversarial training, the classification performance on every class achieve remarkable improvement. 
We can find that A2CNN can make almost every fault classified into the right class.
The high \textit{precision} and \textit{recall} score of A2CNN implies that our model has fewer false alarms and missed alarms. 

Finally, through visualizing the feature maps learned by our model, we explore the inner mechanism of proposed model in fault diagnosis and domain adaptation, and verify that partially untied the layers between the source and target feature extractor is correct.

\section*{References}

\bibliography{zhangbo-20170501}

\end{document}